\documentclass[12pt]{article}
\usepackage{amssymb,latexsym}

\makeatletter \@addtoreset{equation}{section} \makeatother

\begin{document}

\newcommand{\nn}{\nonumber}

\begin{titlepage}

\thispagestyle{empty}

\begin{flushright}
\hfill{preprint number} \\
\hfill{hep-th/yymmnnn}
\end{flushright}

\vspace{20pt}

\begin{center}

{ \Large{\bf Coset Space Dimensional Reduction of
Einstein-Yang-Mills theory }}

\vspace{35pt}

{\bf A.~Chatzistavrakidis}$^{1,2}$,  {\bf P.~Manousselis}$^{2,3}$,
{\bf N.~Prezas}$^4$ {\bf and} {\bf G.~Zoupanos}$^2$ \vspace{20pt}

$^1$ {\it Institute of Nuclear Physics,\\
NCSR  DEMOKRITOS,\\
GR-15310 Athens, Greece}\\
 \vspace{5pt}

$^2${\it Physics Department, National Technical University of Athens, \\
GR-15780 Zografou Campus, Athens, Greece} \\
\vspace{5pt}

$^3${\it Department of Engineering Sciences, University of Patras,\\
GR-26110 Patras, Greece}\\ \vspace{5pt}

$^4${\it CERN PH-TH,\\
1211 Geneva, Switzerland}\\
\vspace{5pt}

\

Email: {\tt cthan@mail.ntua.gr,  pman@central.ntua.gr,
george.zoupanos@cern.ch, nikolaos.prezas@cern.ch}

\vspace{35pt}

{ABSTRACT}
\end{center}
In the present contribution we extend our previous work by
considering the coset space dimensional reduction of
higher-dimensional Einstein--Yang--Mills theories including scalar
fluctuations as well as Kaluza--Klein excitations of the
compactification metric and we describe the gravity-modified rules
for the reduction of non-abelian gauge theories.

\vspace{20pt}

%\end{center}

\vspace{20pt}

\end{titlepage}

\newpage

\baselineskip 6 mm

\section{Introduction}

In the last four decades we have witnessed a revival of interest in
Kaluza--Klein theories, triggered by the realization \cite{Kerner68}
that non-abelian gauge groups appear naturally when one assumes that
the unification takes place in higher dimensions. More specifically,
one typically considers a total space-time manifold that can be
written as a direct product $M^D = M^4 \times B$, where $B$ is a
Riemannian space  with a non-abelian isometry group $S$. The
dimensional reduction of this theory leads to gravity coupled to a
Yang--Mills theory with a gauge group containing $S$ and scalars in
four dimensions. The main advantage of this scenario is the
geometrical unification of gravity with the other interactions and
the natural emergence of the observed non-abelian gauge symmetries.
However, there are problems in the Kaluza-Klein framework.

The most serious obstacle in obtaining a realistic model of the
low-energy interactions is that it is impossible to obtain chiral
fermions in four dimensions \cite{Witten:1981me}. Fortunately, there
is a very interesting resolution to this problem resulting when one
adds Yang--Mills fields to the original gravity action. These gauge
fields can be responsible for a non-trivial background configuration
which could provide chiral fermions to the four-dimensional theory
according to the Atiyah-Hizebruch theorem \cite{Chapline:1982wy}.
Moreover, the system admits a stable classical ground state of the
required form and the relevant mechanism is known as {\it
spontaneous compactification } \cite{Horvath:1977st}. Thus one is
led to introduce Yang--Mills fields in higher dimensions. This
approach is further justified by other popular unification schemes
such as supergravity and heterotic string theory
\cite{Gross:1985fr}.

Gauge fields in the higher-dimensional theory are also welcome from
another point of view, since they can provide a potential
unification of the low-energy gauge interactions as well as of gauge
and Higgs fields. Concerning the latter we should recall that the
celebrated Standard Model (SM) of Elementary Particle Physics, which
had so far outstanding successes in all its confrontations with
experimental results, has also obvious limitations due to the
presence of a plethora of free parameters mostly related to the
ad-hoc introduction of the Higgs and Yukawa sectors in the theory.
The Coset Space Dimensional Reduction (CSDR) \cite{Witten:1976ck,
Forgacs:1979zs, Kapetanakis:1992hf} was suggesting from the
beginning that a unification of the gauge and Higgs sectors can be
achieved using higher dimensions. In the CSDR one assumes that the
form of space-time is $M^D = M^4 \times S/R$ with $S/R$ being a
homogeneous space (obtained as the quotient of the Lie group $S$ by
the Lie subgroup $R$). Then a gauge theory with gauge group $G$
defined on $M^D$ can be dimensionally reduced to $M^4$ in an elegant
way using the symmetries of $S/R$. In particular, the resulting
four-dimensional gauge group is a subgroup of $G$. The
four-dimensional gauge and Higgs fields are simply the surviving
components of the gauge fields of the pure higher-dimensional gauge
theory.

Similarly, when fermions are introduced \cite{Manton:1981es} the
four-dimensional Yukawa and gauge interactions of fermions find also
a unified description in the gauge interactions of the
higher-dimensional theory. The last step in this unified description
in high dimensions is to relate the gauge and fermion fields that
have been introduced. A simple way to achieve that is by demanding
that the higher-dimensional gauge theory is ${\cal N} = 1$
supersymmetric, which requires that the gauge and fermion fields are
members of the same vector supermultiplet. A very welcome additional
input is that heterotic string theory suggests the dimension and the
gauge group of the higher dimensional supersymmetric theory
\cite{Gross:1985fr}. Moreover, ref.~\cite{Manousselis:2005xa} showed
that coset spaces with nearly-K\"ahler geometry yield supersymmetric
solutions of heterotic strings in the presence of fluxes and
condensates. Therefore, the CSDR might be an appropriate reduction
scheme for such compactifications.

The fact that the SM is a chiral theory leads us to consider
$D$-dimensional supersymmetric gauge theories with $D = 4n + 2$
\cite{Chapline:1982wy, Kapetanakis:1992hf}, which include the ten
dimensions suggested by heterotic strings \cite{Gross:1985fr}.
Concerning supersymmetry, the nature of the four-dimensional theory
depends on the nature of the corresponding compact space used to
reduce the higher-dimensional theory. Specifically, the reduction
over CY spaces leads to supersymmetric theories \cite{Gross:1985fr}
in four dimensions, the reduction over symmetric coset spaces leads
to non-supersymmetric theories, while a reduction over non-symmetric
ones leads to softly broken supersymmetric theories
\cite{Manousselis:2001re}.

In the present paper, continuing our recent work on the CSDR of the
bosonic part of a higher-dimensional Einstein--Yang--Mills theory
\cite{Chatzistavrakidis:2007by}, we apply the CSDR to the gravity
sector and describe explicitly the low-energy effective theory. We
emphasize that the latter is characterized by a potential for the
metric moduli. Furthermore, we revisit the CSDR of gauge theories
taking into account the contribution of the dynamical (non-frozen)
gravity background and write down the resulting modified constraints
and effective action.

\section{Geometry of Coset spaces}
To describe the geometry of coset spaces we rely on
refs.~\cite{Castellani:1983tb, Castellani:1999fz}. In the present
section we collect the definitions and results that are useful for
our discussion. On a coset $S/R$ the Maurer-Cartan 1-form is defined
by $e(y)=L^{-1}(y)dL$, where $L(y^{a})$ is a coset representative
and $a= 1 \ldots dim S/R$. It is the analogue of the left-invariant
forms defined on group manifolds and its values are in $Lie(S)$, the
Lie algebra of $S$, i.e.~it can be expanded as
\begin{equation}
e(y) = e^{A}Q_{A}= e^{a}Q_{a} + e^{i}Q_{i},
\end{equation}
where $A$ is a group index, $a$ is a coset index and $i$ is an
$R$-index. $e^{a}$ is the coframe and $e^{i}$ is the $R$-connection.
The exterior derivative of the Maurer-Cartan 1-form is
\begin{equation}\label{2.4}
de^{A} = - \frac{1}{2}f^{A}_{ \ \ BC} e^{B} \wedge e^{C}.
\end{equation}
Eq.~(\ref{2.4}) can be expanded  as
\begin{eqnarray}\label{first}
de^{a} = - \frac{1}{2}f^{a}_{ \ \ bc}e^{b} \wedge e^{c} -
f^{a}_{ \ \ bi}e^{b} \wedge e^{i}, \nonumber \\
de^{i} = -\frac{1}{2} f^{i}_{ \ \ ab}e^{a} \wedge e^{b} -
\frac{1}{2}f^{i}_{ \ \ jk}e^{j} \wedge e^{k}.
\end{eqnarray}
The commutation relations obeyed by the generators of $S$ are
\begin{eqnarray}\label{2.6}
\left[ Q_{i}, Q_{j} \right] &=& f_{ij}^{ \ \ k} Q_{k}, \nonumber \\
\left[Q_{i}, Q_{a}\right] &=& f_{ia}^{ \ \ b}Q_{b}, \nonumber \\
\left[Q_{a},Q_{b}\right]&=& f_{ab}^{ \ \ c} Q_{c} + f_{ab}^{\ \ i}
Q_{i}.
\end{eqnarray}

We assume (for reasons analyzed in detail in
ref.~\cite{Castellani:1983tb}) that the coset is reductive,
i.e.~$f_{bi}^{ \ \ j}=0$. The normalizer $N(R)$ of $R$ in $S$ is
defined as follows
\begin{equation}
N = \{ s \in S, \ \ \  sRs^{-1} \subset R \}.
\end{equation}
Since $R$ is normal in $N(R)$ the quotient $N(R)/R$ is a group. The
generators $Q_{a}$ split into two sets $Q_{\hat{a}},Q_{\bar{a}}$
with $Q_{\hat{a}}$ forming a group which is isomorphic to $N(R)/R$.
Then the Lie algebra of $S$ decomposes as
$$S = R + K + L,$$ with
\begin{equation}\label{decomp1}
\left[K,K \right] \subset K, \ \ \ \left[K, R\right]=0, \ \ \
\left[K, L \right] \subset L, \ \ \ \left[L, R\right] \subset L, \ \
\  \left[L,L\right] = L+R.
\end{equation}
Accordingly, the commutation relations (\ref{2.6}) split as
\begin{eqnarray}\label{decomp2}
\left[Q_{\hat{a}}, Q_{\hat{b}} \right] &=& f_{\hat{a} \hat{b}}^{ \ \
\hat{c}}Q_{\hat{c}},~~~\left[Q_{i}, Q_{\hat{a}} \right] =
0,~~~\left[Q_{\hat{a}}, Q_{\bar{a}} \right] = f_{\hat{a} \bar{a}}^{
\ \
\bar{b}}Q_{\bar{b}},\nonumber \\
\left[Q_{i}, Q_{\bar{a}} \right] &=& f_{i \bar{a}}^{ \ \
\bar{b}}Q_{\bar{b}},~~~ \left[Q_{\bar{a}}, Q_{\bar{b}}\right] =
f_{\bar{a} \bar{b}}^{\ \ \bar{c}}Q_{\bar{c}}+ f_{\bar{a} \bar{b}}^{
\ \ i}Q_{i}.
\end{eqnarray}
Eq.~(\ref{first}) is then further decomposed to
\begin{eqnarray}
de^{\hat{a}} &=& - \frac{1}{2}f^{\hat{a}}_{ \ \ \hat{b} \hat{c}}
e^{\hat{b}} \wedge e^{\hat{c}}, \nonumber \\
de^{\bar{a}}&=& -\frac{1}{2}f^{\bar{a}}_{ \ \ \bar{b}
\bar{c}}e^{\bar{b}} \wedge e^{\bar{c}} - f^{\bar{a}}_{ \ \ \hat{b}
\bar{c}}e^{\hat{b}} \wedge e^{\bar{c}}-f^{\bar{a}}_{ \ \ \bar{b}
i}e^{\bar{b}}
\wedge e^{i}, \nonumber \\
de^{i} &=& -\frac{1}{2}f^{i}_{ \ \ \bar{b} \bar{c}}e^{\bar{b}}
\wedge e^{\bar{c}}-\frac{1}{2}f^{i}_{ \ \ j k}e^{j} \wedge e^{k}.
\end{eqnarray}

An $S$-invariant metric on $S/R$ is
\begin{equation}\label{Smetric}
g_{\alpha\beta}(y) = \delta_{a b}e^{a}_{\alpha}(y)e^{b}_{\beta}(y).
\end{equation}
Using the metric (\ref{Smetric}) the following useful identities can
be proved
\begin{eqnarray}\label{ident}
&&e^{a} \wedge \ast_{d} e^{b} = \delta^{ab} vol_{d},\\
%&&e^{a} \wedge \ast_{d} (e^{b} \wedge e^{c}) = \delta^{ac} \ast_{d}
%e^{b} - \delta^{ab} \ast_{d} e^{c},\\
&&(e^a \wedge e^b) \wedge *_{d} (e^c \wedge e^d) = \delta^{ab}_{cd}
vol_{d},\\
&&(e^a \wedge e^b \wedge e^c) \wedge *_{d} (e^d \wedge e^e \wedge
e^f) = \delta^{abc}_{def} vol_{d}.
\end{eqnarray}
where $\ast_{d}$ is the Hodge duality operator on a $d$-dimensional
coset. The Killing vectors associated with the left-isometry group
$S$ are
\begin{equation}
K_{A}^{\alpha} = D_{A}^{a}e_{a}^{\alpha},
\end{equation}
where $e_{a}^{\alpha}$ is the inverse vielbein and $D_{A}^{B}(s)$ is
a matrix in the adjoint representation of $S$. The coset $S/R$ also
posses a right-isometry group which is $N(R)/R$. The relevant
Killing vectors are
\begin{eqnarray}
%\tilde{K}_{i}^{\alpha} = 0, \nonumber \\
\tilde{K}_{\hat{a}}^{\alpha} = e_{\hat{a}}^{\alpha},
\end{eqnarray}
where $\hat{a} = 1 \ldots dim N(R)/R$ and $e_{\hat{a}}^{\alpha}$ is
the inverse vielbein.

\section{The Coset Space Dimensional Reduction}

In the present section we present a brief reminder of the Coset
Space Dimensional Reduction scheme. The CSDR  of a multidimensional
gauge field $\hat{A}$ on a coset $S/R$ is a truncation described by
a generalized invariance condition
\begin{equation}
{\cal L}_{X^{I}}\hat{A} = DW_{I},
\end{equation}
where $W_{I}$ is a parameter of a gauge transformation associated
with the Killing vector $X_{I}$ of $S/R$. The relevant invariance
condition for the reduction of the metric is
\begin{equation}\label{InvarianceCondition}
{\cal L}_{X^{I}}g_{MN} =0.
\end{equation}
The generalized invariance condition
\begin{equation}\label{21}
{\cal L}_{X^{I}}\hat{A} = i_{X^{I}}d\hat{A} + di_{X^{I}}\hat{A} =
DW_{I} = dW_{I} + [\hat{A},W_{I}],
\end{equation}
together with the consistency condition
\begin{equation}\label{22}
\left[{\cal L}_{X^{I}}, {\cal L}_{X^{J}} \right] = {\cal
L}_{[X^{I},X^{J}]},
\end{equation}
impose constraints on the gauge field. The detailed analysis of the
constraints (\ref{21}) and (\ref{22}), given in
refs.\cite{Forgacs:1979zs,Kapetanakis:1992hf}, provides us with the
four-dimensional unconstrained fields as well as with the gauge
invariance that remains in the theory after dimensional reduction.

Instead, we may use the following  ansatz for the gauge fields,
which was shown in \cite{Chatzistavrakidis:2007by} to be equivalent
to the CSDR ansatz and it is similar to the Scherk-Schwartz
reduction ansatz:
\begin{equation}
\hat{A}^{\tilde{I}}(x,y) = A^{\tilde{I}}(x) +
\chi^{\tilde{I}}_{\alpha}(x,y)dy^{\alpha},
\end{equation}
where
\begin{equation}
\chi^{\tilde{I}}_{\alpha}(x,y) =
\phi^{\tilde{I}}_{A}(x)e^{A}_{\alpha}(y).
\end{equation}
The objects $\phi_{A}(x)$, which take values in the Lie algebra of
$G$, are coordinate scalars in four dimensions and they can be
identified with Higgs fields.
%{\it CSDR of Yang-Mills theory}

\section{Gravity and CSDR}

Usually one studies higher-dimensional gauge theories and constructs
four-dimensional unified models, in a frozen gravity background,
i.e., the internal metric is of the form (\ref{Smetric}).  In this
section we search for gravity backgrounds consistent with CSDR in
the sense of eq.~(\ref{InvarianceCondition}) but including
fluctuations of the metric \cite{Palla:1983re, Coquereaux:1984ca,
Pilch:1984ur, Schellekens:1984dm}. We begin by examining a
$D$-dimensional Einstein--Yang--Mills Lagrangian
\begin{equation}
{\cal L} = \hat{R} \ast_{D} {\bf 1} - \frac{1}{2}Tr\hat{F}_{(2)}
\wedge \ast_{D} \hat{F}_{(2)} - \hat{\lambda}_{(D)} \ast_{D} {\bf
1},
\end{equation}
where $\hat{F}_{(2)}=d\hat{A}_{(1)} + \hat{A}_{(1)} \wedge
\hat{A}_{(1)}$ is a gauge field with values in the Lie algebra of a
group $G$, $\hat{R}$ is the curvature scalar and
$\hat{\lambda}_{(D)}$ is the cosmological constant in
$D$-dimensions. A general ansatz for the metric is
\begin{equation}\label{ouransatz}
d\hat{s}_{(D)}^2 = ds_{(4)}^2 + h_{\alpha\beta}(x,y)(dy^\alpha -
{\cal A}^\alpha(x,y))(dy^\beta - {\cal A}^\beta(x,y)),
\end{equation}
where $A^\alpha$ is the Kaluza-Klein gauge field
\begin{equation}
{\cal A}^\alpha(x,y) = {\cal A}^I(x) K^{\alpha}_{(I)}(y), \ \ {\cal
A}^{I}(x) = {\cal A}^{I}_{\mu}(x) dx^{\mu},
\end{equation}
and $ K_{(I)}(y) = K^{\alpha}_{(I)}(y) \frac{\partial}{\partial
y^{\alpha}}$ are at most the $dim S + dim( N(R)/R)$ Killing vectors
of the coset $S/R$ or an appropriate subset. A well known problem
with coset reductions is that we cannot consistently allow
Kaluza--Klein gauge fields from the full isometry group $S$ of the
coset $S/R$ to survive.

According to refs \cite{Coquereaux:1986zf}, \cite{Cho:1987jf} the
correct ansatz leading to a consistent truncation of the theory is
to consider Kaluza-Klein gauge fields belonging to the $N(R)/R$ part
of the isometry group $S/R$
\begin{equation}
{\cal A}^{\alpha}(x,y) = {\cal A}^{\hat{a}}(x) \tilde{K}^{\alpha}_{
\ \hat{a}}(y).
\end{equation}
Now in  the ansatz (\ref{ouransatz}) we have
\begin{equation}
\eta^{\hat{a}} = e^{\hat{a}}_{\alpha}(dy^{\alpha} - A^{\hat{b}}(x)
\tilde{K}^{\alpha}_{ \ \hat{b}}(y)) = e^{\hat{a}} - A^{\hat{a}}(x),
\end{equation}
given that
\begin{equation}
e^{\hat{a}}_{\alpha}\tilde{K}^{\alpha}_{\hat{b}} =
\delta^{\hat{a}}_{\hat{b}},
\end{equation}
with $\tilde{K}^{\alpha}_{\hat{b}}$ being the Killing vectors of the
right isometries $N(R)/R$. The rest of the $1$-forms are
\begin{equation}
\eta^{\bar{a}} = e^{\bar{a}}, \ \ \ e^{i} = e^{i}_{a}e^{a}.
\end{equation}
For $\eta^{\hat{a}}$ we find that
\begin{equation}\label{937}
D\eta^{\hat{a}}\equiv d\eta^{\hat{a}} + f^{\hat{a}}_{\ \ \hat{b}
\hat{c}} {\cal A}^{\hat{b}} \wedge \eta^{\hat{c}} = - {\cal
F}^{\hat{a}} - \frac{1}{2}f^{\hat{a}}_{\ \ \hat{b}
\hat{c}}\eta^{\hat{b}} \wedge \eta^{\hat{c}},
\end{equation}
where ${\cal F}^{\hat{b}}$ is the field strength of the
Kaluza--Klein gauge field ${\cal A}^{\hat{b}}$ defined by
\begin{equation}
{\cal F}^{\hat{b}} = d{\cal A}^{\hat{b}} + \frac{1}{2}
f^{\hat{b}}_{\ \ \hat{c} \hat{d} }{\cal A}^{\hat{c}} \wedge {\cal
A}^{\hat{d}}.
\end{equation}

Now the metric ansatz for a general $S$-invariant metric takes the
form
\begin{equation}
d\hat{s}_{(D)}^2 = e^{2\alpha\phi(x)} \eta_{mn} e^{m} e^{n} +
e^{2\beta\phi(x)}\gamma_{ab}(x)\eta^{a}\eta^{b},
\end{equation}
from which we read the vielbeins (the notation is  close to that one
used in ref.~\cite{Cvetic:2003jy}):
\begin{equation}
\hat{e}^m= e^{\alpha\phi}e^m, \ \ \ \hat{e}^{a} =
e^{\beta\phi}\Phi^{a}_{b}(x)\eta^{b},
\end{equation}
with
\begin{equation}
\gamma_{cd}(x) = \delta_{ab} \Phi^{a}_{c}(x) \Phi^{b}_{d}(x).
\end{equation}
$\Phi$ is a matrix of unit determinant so there exists a set
$(\Phi^{-1})^{b}_{a}$ of fields satisfying
\begin{equation}
(\Phi^{-1})^{c}_{a}(\Phi^{-1})^{d}_{b}\gamma_{cd} = \delta_{ab}.
\end{equation}
Next we calculate the exterior derivatives of the vielbeins
\begin{equation}
d\hat{e}^{m}= - \omega^{m}_{\ \ n} \wedge \hat{e}^{n}-\alpha
e^{-\alpha\phi}\partial_{\nu}\phi\hat{e}^m\wedge\hat{e}^n,
\end{equation}
\begin{eqnarray}
d\hat{e}^{a} = -\tilde{f}^{a}_{ \ i b} e^i \wedge \hat{e}^{b}+
e^{-\alpha\phi}D_{bn}^{\ \ \ a} \hat{e}^{nb} +\beta
e^{-\alpha\phi}\partial_m\phi\hat{e}^{ma}-
\frac{1}{2}e^{(\beta-2\alpha)\phi}{\cal F}^{a}_{ \ \ mn}
\hat{e}^{mn} - \frac{1}{2}\tilde{f}^{a}_{ \ bc} \hat{e}^{bc},
\end{eqnarray}
where
\begin{equation}
e^{ab} \equiv e^{a} \wedge e^{b}, \ \ \ \ {\cal F}^{a}_{mn} \equiv
\Phi^{a}_{\hat{a}}{\cal F}^{\hat{a}}_{mn},
\end{equation}
and
\begin{eqnarray}\label{fitilde}
&&\tilde{f}^{a}_{ \ ib} = \Phi^{a}_{c}(\Phi^{-1})^{d}_{b}f^{c}_{ \
id}, \ \ \  D_{bn}^{\ \ \ a} =(\Phi^{-1})_{b}^{ \ c}D_{n}\Phi^{a}_{
\ c},\nonumber
\\&&\tilde{f}^{a}_{ \ bc} = \Phi^{a}_{
\ d}(\Phi^{-1})_{b}^{ \ e}(\Phi^{-1})_{c}^{ \ f}f^{d}_{ \ \ ef}.
\end{eqnarray}

Subsequently  we compute  the spin connections
\begin{equation}\label{spcon1}
\hat{\omega}_{mn} = \omega_{mn}  + \frac{1}{2}
e^{(\beta-2\alpha)\phi}{\cal F}^{a}_{ \ \ mn} \hat{e}^{a}+\alpha
e^{-\alpha\phi}(\partial_n\phi\eta_{ml}\hat{e}^l-\partial_m\phi\eta_{nl}\hat{e}^l),
\end{equation}
\begin{equation}\label{spcon2}
\hat{\omega}_{ma} = - e^{-\alpha\phi}P_{mab} \hat{e}^{b}-\beta
e^{-\alpha\phi}\partial_m\phi\hat{e}^a+
\frac{1}{2}e^{(\beta-2\alpha)\phi}{\cal F}_{aml} \hat{e}^{l},
\end{equation}
\begin{equation}\label{spcon3}
\hat{\omega}_{ab}= -\tilde{f}_{iab}e^{i} + e^{-\alpha\phi} Q_{mab}
\hat{e}^{m} + e^{-\beta\phi}\tilde{C}_{cab} \hat{e}^{c},
\end{equation}
where
\begin{eqnarray}{\label{orismoi}}
&&\tilde{C}_{cab} = \frac{1}{2} (\tilde{f}^{c}_{ \ \ ab} +
\tilde{f}^{b}_{
\ \ ac} - \tilde{f}^{a}_{ \ \ bc}), \nonumber \\
&&P_{mab} = \frac{1}{2}[(\Phi^{-1})^{c}_{a}D_{m}\Phi^{b}_{c} +
(\Phi^{-1})^{c}_{b}D_{m}\Phi^{a}_{c}], \nonumber \\
&&Q_{mab} = \frac{1}{2}[(\Phi^{-1})^{c}_{a}D_{m}\Phi^{b}_{c} -
(\Phi^{-1})^{c}_{b}D_{m}\Phi^{a}_{c}], \nonumber \\
&&D_{m}\Phi^{a}_{d} = \partial_{m}\Phi^{a}_{d} - f^{c}_{ \ \ \ d
\hat{b}} {\cal A}_{m}^{\hat{b}} \Phi^{a}_{c}.
\end{eqnarray}

It is well-known that the curvature scalar of the gravitational
Lagrangian can be written as
\begin{equation}
\hat{R} *_{D} {\bf 1} = \hat{\Theta}_{AB} \wedge *_{D}(\hat{e}^{a}
\wedge \hat{e}^{B}),
\end{equation}
where $A=m,a$ and $\Theta_{AB}$ are the curvature 2-forms calculated
from eqs.~(\ref{spcon1}), (\ref{spcon2}) and  (\ref{spcon3}). Then
the Lagrangian is reduced to four dimensions provided we impose the
following constraints
\begin{eqnarray}\label{constraints}
&&-\frac{1}{2} \tilde{f}^{a}_{ib}f^{i}_{jk} +
\tilde{f}^{a}_{jc}\tilde{f}^{c}_{kb}=0, \nonumber \\
&&-\tilde{C}^{a}_{cb}\tilde{f}^{c}_{id}+\tilde{C}^{a}_{dc}\tilde{f}^{c}_{ib}-\tilde{C}^{c}_{db}\tilde{f}^{a}_{ic}=0.
\end{eqnarray}
The constraints (\ref{constraints}) can be shown to be satisfied
using the Jacobi identities and the invariance of the metric.

Finally, we can write down the reduced Lagrangian in the form
\begin{eqnarray}\label{redlang}
\mathcal{L}&=&e^{4\alpha+d\beta}\Big(e^{-2\alpha\phi}R*\mathbf{1}-e^{-2\alpha\phi}*P_{ab}\wedge
P_{ab}-\frac{1}{2}e^{2(\beta-2\alpha)\phi}\gamma_{ab}*\mathcal{F}^a\wedge
\mathcal{F}^b
\nonumber\\&+&e^{-2\alpha\phi}((3\alpha+d\beta)^2-(3\alpha^2+d\beta^2))*d\phi\wedge
d\phi
\nonumber\\&-&\frac{1}{4}e^{-2\beta\phi}(\gamma_{ab}\gamma^{cd}\gamma^{ef}f^{a}_{
\ ce}f^{b}_{ \
 df}+2\gamma^{ab}f^{c}_{ \ da}f^{d}_{ \ cb}+4e^{2\beta\phi}\gamma^{ab}f_{iac}f^{ic}_{b} )*\mathbf{1} + \lambda_{D} *\mathbf{1}\Big).
\end{eqnarray}
In order to obtain the correct kinetic terms in four dimensions we
should  choose
\begin{equation}
\alpha = -\sqrt{\frac{d}{4d+8}},~~~~~ \beta = -\frac{2\alpha}{d}.
\end{equation}
To the set of the imposed constraints we should add that the
condition that $\Phi$ is a matrix of unit determinant and that the
structure constants of $S$ are traceless and fully antisymmetric.
The final form of the reduced Lagrangian is
\begin{equation}\label{redlang1}
\mathcal{L}=R*\mathbf{1}-*P_{ab}\wedge
P_{ab}-\frac{1}{2}e^{2(\beta-\alpha)\phi}\gamma_{\hat{a}
\hat{b}}*\mathcal{F}^{\hat{a}} \wedge
\mathcal{F}^{\hat{b}}-\frac{1}{2}*d\phi\wedge d\phi-V(\phi),
\end{equation}
where the potential for the metric moduli fields reads
\begin{equation}\label{pot}
V=\frac{1}{4}e^{2(\alpha-\beta)\phi}(\gamma_{ab}\gamma^{cd}\gamma^{ef}f^{a}_{
\ ce}f^{b}_{ \
 df}+2\gamma^{ab}f^{c}_{ \ da}f^{d}_{ \ cb}+4e^{2\beta\phi}\gamma^{ab}f_{iac}f^{ic}_{b}-4 e^{2\beta\phi} \lambda_{D} )*\mathbf{1} .
\end{equation}
Note that the first two terms in eq.~(\ref{pot}) have a non-zero
contribution only in the case of non-symmetric coset spaces.

\section{Reduction of the Gauge Sector: Gravity Modification of the  CSDR Rules}

In this section we reduce the Yang--Mills Lagrangian in the presence
of fluctuating gravity. The ansatz for the higher dimensional gauge
field is
\begin{equation}
\hat{A}^{\tilde{I}}=A^{\tilde{I}} + \phi^{\tilde{I}}_{A}\eta^{A},
\end{equation}
where $$\eta^{\hat{a}} = e^{\hat{a}} - {\cal A}^{\hat{a}}, \ \
\eta^{\bar{a}} = e^{\bar{a}}, \ \ \eta^{i} = e^{i}=e^{i}_{a}e^{a},
$$ and $\tilde{I}$ is a gauge group index. Calculating the field strength
\begin{equation}
\hat{F} = \hat{d}\hat{A}^{\tilde{I}} + \frac{1}{2}f^{\tilde{I}}_{ \
\ \tilde{J} \tilde{K}} \hat{A}^{\tilde{J}} \wedge
\hat{A}^{\tilde{K}},
\end{equation}
we find
\begin{equation}\label{Pans}
\hat{F}^{\tilde{I}} = (F^{\tilde{I}} - {\cal
F}^{\hat{a}}\phi_{\hat{a}}^{\tilde{I}}) + D\phi^{\tilde{I}}_{A}
\wedge \eta^{A} - \frac{1}{2}F^{\tilde{I}}_{AB} \eta^{A} \wedge
\eta^{B},
\end{equation}
where ${\cal F}^{\hat{a}}$ is the KK gauge field and
\begin{equation}
F^{\tilde{I}} = dA^{\tilde{I}} + \frac{1}{2}f^{\tilde{I}}_{ \ \
\tilde{J} \tilde{K}} A^{\tilde{J}} \wedge A^{\tilde{K}},
\end{equation}
with
\begin{equation}
F^{\tilde{I}}_{AB} = f^{C}_{ \ \ AB}\phi^{\tilde{I}}_{C}-
f^{\tilde{I}}_{ \ \ \tilde{J} \tilde{K}}\phi^{\tilde{J}}_{A}
\phi^{\tilde{K}}_{B},
\end{equation}
and
\begin{equation}
D\phi^{\tilde{I}}_{A}= d\phi^{\tilde{I}}_{A} + f^{C}_{ \ \ AB}{\cal
A}^{B} \phi^{\tilde{I}}_{C} + f^{\tilde{I}}_{ \ \ \tilde{J}
\tilde{K}}A^{J} \phi^{\tilde{K}}_{A}.
\end{equation}

To reduce the higher dimensional Yang--Mills Lagrangian we dualize
eq.~(\ref{Pans}) to
\begin{equation}\label{DualPans}
\ast_{D}\hat{F}^{\tilde{I}} =  \ast_{4}(F^{\tilde{I}} - {\cal
F}^{\hat{a}}\phi^{\tilde{I}}_{\hat{a}})\wedge vol_{d} +
e^{\alpha\phi-\beta\phi} \ast_{4}D\phi_{A}^{\tilde{I}} \wedge
\ast_{d}\tilde{\eta}^{A} - \frac{1}{2} e^{2\alpha\phi-2\beta\phi}
F^{\tilde{I}}_{AB} vol_{4} \wedge \ast_{d}(\tilde{\eta}^{A} \wedge
\tilde{\eta}^{B}),
\end{equation}
and insert everything in
$$ {\cal L} = -\frac{1}{2} Tr \hat{F} \wedge \ast_{D}\hat{F}.$$
The result is
\begin{eqnarray} \label{KKYM}
{\cal L} &=& -\frac{1}{2} e^{-2\alpha\phi}(F^{\tilde{I}} - {\cal
F}^{\hat{a}}\phi_{\hat{a}}^{\tilde{I}}) \wedge
\ast_{4}(F^{\tilde{I}} - {\cal
F}^{\hat{a}}\phi^{\tilde{I}}_{\hat{a}}) \wedge vol_{d} - \frac{1}{2}
e^{-2\beta\phi} D\phi^{\tilde{I}}_{A} \wedge
\ast_{4}D\phi_{B}^{\tilde{I}} \wedge \tilde{\eta}^{A} \wedge
\ast_{d} \tilde{\eta}^{B} \nonumber
\\ &+& \frac{1}{4}e^{2\alpha\phi-4\beta\phi} F_{AB}F_{CD} vol_{4} \wedge \tilde{\eta}^{A} \wedge
\tilde{\eta}^{B} \wedge \ast_{d}(\tilde{\eta}^{C} \wedge
\tilde{\eta}^{D}),
\end{eqnarray}
where
\begin{equation}
\tilde{\eta^{a}} = (\Phi^{-1})^{a}_{b}\eta^{b}, \ \ \
\tilde{\eta^{i}} = e^{i}_{a}(\Phi^{-1})^{a}_{b}\eta^{b}.
\end{equation}
To reduce eq.~(\ref{KKYM}) we must impose the constraints
\begin{equation}\label{newcon1}
D\phi^{\tilde{I}}_{i} =0, \ \ F^{\tilde{I}}_{ij} =
F^{\tilde{I}}_{aj} = 0,
\end{equation}
and
\begin{equation}
D\phi_{\hat{a}} = F_{\hat{a} \hat{b}} = F_{\hat{a} \bar{b}} =
F_{\hat{a} i} =0.
\end{equation}

The first set is the  usual CSDR constraints described in detail at
various places (e.g.\cite{Chatzistavrakidis:2007by}). We concentrate
on the gravity-induced second set. From the condition
$$ F^{\tilde{I}}_{\hat{a} \hat{b}} = f^{\hat{c}}_{ \ \ \hat{a}
\hat{b}}\phi^{\tilde{I}}_{\hat{c}} - [\phi_{\hat{a}},
\phi_{\hat{b}}]^{\tilde{I}} = 0,$$ we conclude that $\phi_{\hat{a}}$
are the generators of an $N(R)/R$ subgroup of $H$ (remember that $R$
has no $N(R)/R$ subgroup and $H$ is the centralizer of the embedding
of $R$ on $G$, the higher dimensional gauge group). We conclude also
that
\begin{equation}\label{condition1}
f^{\hat{c}}_{ \ \ \hat{a} \hat{b}}\phi^{\tilde{I}}_{\hat{c}}
=f^{\tilde{I}}_{ \ \ \tilde{J} \tilde{K}} \phi_{\hat{a}}^{\tilde{J}}
\phi^{\tilde{K}}_{\hat{b}}.
\end{equation}
Given the condition (\ref{condition1}) the constraint
$D\phi_{\hat{a}}^{\tilde{I}} = 0$ yields ($\phi_{\hat{a}}$ is
constant)
\begin{equation}\label{condition2}
f^{\tilde{I}}_{ \ \ \tilde{J} \tilde{K}}\phi^{\tilde{J}}_{\hat{a}}
\phi_{\hat{b}}^{\tilde{K}} {\cal A}^{\hat{b}} +f^{\tilde{I}}_{ \ \
\tilde{J} \tilde{K}}A^{\tilde{J}} \phi^{\tilde{K}}_{\hat{b}} = 0.
\end{equation}
Eq.~(\ref{condition2}) determines the gauge field belonging to the
$N(R)/R$ part of $H$ in terms of the Kaluza--Klein gauge fields
\begin{equation}
A^{\tilde{I}} = {\cal A}^{\hat{b}} \phi_{\hat{b}}^{\tilde{I}}.
\end{equation}
Calculating the corresponding field strength we find
\begin{equation}
F^{\tilde{I}} = d{\cal A}^{\hat{a}} \phi_{\hat{a}}^{\tilde{I}} +
\frac{1}{2}f^{\hat{c}}_{ \ \ \hat{a} \hat{b}}{\cal A}^{\hat{a}}
\wedge {\cal A}^{\hat{b}} \phi^{\tilde{I}}_{\hat{c}} = {\cal
F}^{\hat{a}} \phi_{\hat{a}}^{\tilde{I}}.
\end{equation}
This is exactly the term subtracted from $F^{\tilde{I}}$ in
eq.~(\ref{Pans}), thus leaving a surviving gauge group $K$ obtained
from the decompositions
$$ G \supset R \times H$$
and
$$H \supset (N(R)/R) \times K.$$
The constraint $$F_{\hat{a}i} = [\phi_{\hat{a}}, \phi_{i}] =0 $$ is
satisfied trivially while the representations in which the scalars
$\phi_{\hat{a}}$ belong are determined by
\begin{eqnarray}
F_{\hat{a} \bar{b}} &=& f^{\bar{c}}_{ \ \ \hat{a}
\bar{b}}\phi_{\bar{c}} - [ \phi_{\hat{a}},
\phi_{\bar{b}}] =0, \\
F_{\bar{a}i}&=& f^{\hat{c}}_{ \ \ \bar{a} i}\phi_{\hat{c}} - [
\phi_{\bar{a}}, \phi_{i}] =0.
\end{eqnarray}
These constraints are solved by considering the following
decompositions of $S$ and $G$
\begin{eqnarray}
S &\supset& R \times (N(R)/R), \nonumber \\
adS &=& adR + ad N(R)/R + \sum (r_{i}, n_{i})
\end{eqnarray}
and
\begin{eqnarray}
G &\supset& R \times  (N(R)/R) \times K, \nonumber \\
adG &=& (adR,1,1) + (1,ad N(R)/R ,1) + (1,1,adK) + \sum (l_{i},
m_{i}, k_{i}).
\end{eqnarray}
As in the pure Yang--Mills case  there is a $k_{i}$ multiplet of
scalar fields surviving when $(r_{i}, n_{i}) = (l_{i}, m_{i})$.

Collecting the various terms  we obtain the Lagrangian
\begin{equation}
{\cal L} =-\frac{1}{2}e^{-2\alpha\phi} F^{\tilde{I}} \wedge \ast_{4}
F^{\tilde{I}} \wedge vol_{d} -
\frac{1}{2}e^{-2\beta\phi}\gamma^{\bar{a} \bar{b}}
D\phi_{\bar{a}}^{\tilde{I}} \wedge
\ast_{4}D\phi_{\bar{b}}^{\tilde{I}} \wedge vol_{d} +
\frac{1}{4}e^{2\alpha\phi-4\beta\phi}\gamma^{\bar{a} \bar{c}}
\gamma^{\bar{b} \bar{d}} F_{\bar{a} \bar{b}}F_{\bar{c}
\bar{d}}vol_{4} \wedge vol_{d},
\end{equation}
with gauge group $K$ and scalars in specific representations of $K$
subject to the potential
\begin{equation}
V_{gt} = - \frac{1}{4}e^{(2\alpha\phi-4\beta\phi)} \gamma^{\bar{a}
\bar{c}} \gamma^{\bar{b} \bar{d}} F_{\bar{a} \bar{b}}F_{\bar{c}
\bar{d}}.
\end{equation}

\section{Conclusions}

We have studied higher-dimensional Einstein--Yang--Mills theories
and examined their Coset Space Dimensional Reduction using an
approach similar to that of ref.~\cite{Cvetic:2003jy} and  combined
with the method of Coset Space Dimensional Reduction of gauge
theories introduced in ref.~\cite{Forgacs:1979zs}. We found that the
expected four-dimensional gauge theory coming from CSDR
considerations with frozen metric is indeed enhanced by the
Kaluza--Klein modes of the metric. However, the emergence of the
full isometry of the coset as a part of the four-dimensional gauge
group is not permitted. In addition, we showed how the
four-dimensional potential is modified from the new scalar fields in
the case of non-symmetric coset spaces.

Ref.~\cite{Manousselis:2005xa} uncovered supersymmetric vacua of
heterotic supergravity (with fluxes and condensates) of the form
$M_{1,3} \times S/R$, with $S/R$ being  a homogeneous
nearly-K\"ahler manifold. It would be interesting to perform
explicitly the reduction on these manifolds using the scheme
developed in this work and compare it with the approach of
\cite{House:2005yc} for reduction on $SU(3)$ structure manifolds.

\section*{Acknowledgements}
This work is supported by the EPEAEK programmes Pythagoras (
co-founded by the European Union (75 \%) and the Hellenic State (25
\%) ) and in part by the European Commission under the Research and
Training Network contract MRTN-CT-2004-503369; PM is supported by
the Hellenic State Scholarship Foundation (I.K.Y), by the programme
Pythagoras I (89194) and by the NTUA programme for fundamental
research ``K. Caratheodory".

\end{document}